\documentclass{appolb}
\usepackage{epsfig}
\usepackage{amsmath,amssymb,bm}

\newcommand{\be}{\begin{equation}}
\newcommand{\ee}{\end{equation}}
\newcommand{\bea}{\begin{eqnarray}}
\newcommand{\eea}{\end{eqnarray}}

\newcommand{\sun}{{\odot}}

\begin{document}
\title{\bf Two solar-mass compact stars: structure, composition, and cooling%
\thanks{Lecture
presented at ``Three Days on Quarkyonic Island'', HIC for FAIR
workshop and XXVIII Max Born Symposium, Wroc\l aw, 19-21 May 2011.
}%
}
\author{Armen Sedrakian\thanks{Also 
Department of Physics, 
Yerevan State University, Armenia}
\address{
Institute for Theoretical Physics,
J. W. Goethe-University,\\D-60438 Frankfurt am Main, Germany
}
}

\maketitle
\begin{abstract}
  I discuss the structure and composition of massive (two solar-mass)
  neutron stars containing hypernuclear and deconfined quark matter in
  color superconducting states. 
  Stable configurations featuring such matter are obtained if the
  equation of state of hadronic matter is stiff above the saturation
  density, the transition to quark matter takes place at a few times
  the nuclear saturation density, and the repulsive vector
  interactions in quark matter are substantial. I also discuss 
  our recent progress in understanding the cooling of massive compact
  stars with color superconducting quark cores.
\end{abstract}
  
\section{Introduction}
\label{sec:Intro}

The heavy-ion collision experiments and compact stars provide mutually
complementary channels to address some of the outstanding challenges
of the modern particle and nuclear physics. While the
energy densities achieved in these experiment are overlapping with the
range that must exist in neutron (or more generally compact) stars,
their astrophysical studies still provide a window on the properties
of dense matter that may be difficult or even impossible to extract
otherwise.

The masses of neutron stars are the most sensitive among their
integral parameters to the equation of state (hereafter EOS) at high
densities. Therefore, pulsar mass measurements provide one of the key
experimental constraints on the theory of ultra-dense
matter~(e.g. Refs.~\cite{weber_book,Sedrakian:2006mq}).  The masses
measured in the pulsar binaries are clustered around the value
1.4~$M_{\sun}$ and have been consider as ``canonical'' for a long
time.  However, in recent years mounting evidence emerged in favor of
substantially heavier neutron stars with $M\le 2M_{\sun}$.  In
particular, the recent discovery of a compact star with a mass of
1.97~$M_{\sun}$ measured through the Shapiro delay provides an
observationally ``clean'' lower bound on the maximum mass of a compact
star~\cite{Demorset:2010}.

On the theoretical side it is now well-established that the emergence
of new degrees of freedom at high densities softens the EOS of
matter. For example, allowing for hyperons can reduce the maximum mass
of a sequence of compact stars below the canonical mass of
1.4~$M_{\sun}$. A similar reduction may occur if a deconfinement to
quark matter takes place, although the softening of the EOS in this
case is less dramatic. Thus, the observation of 2$M_{\sun}$ mass
neutron star is evidence that the ultra-dense matter in neutron stars
cannot be soft, \ie, agents that will substantially soften the EOS are
potentially excluded. Thus, one of the outstanding challenges in the
theory of compact objects is to exploit fully the consequences of this
recent observation and, for example, to provide new EOSs that are
capable to produce compact objects as massive as the millisecond
pulsar J1614-2230.

The cooling of neutron stars provides another channel on the
properties of dense matter, which is sensitive to the {\it
  composition} of matter and weak interactions therein. The cooling of
compact stars can be divided roughly in the following phases. After
the initial non-isothermal phase of rapid cooling from
temperatures $T \sim 50$ MeV down to 0.1 MeV, a neutron star settles
in a thermal quasi-equilibrium state which evolves slowly over the
time scales $10^3-10^5$ yr down to temperatures $T \sim 0.01$
MeV~\cite{Yakovlev:2007vs,Page:2005fq}.  In this latter phase the core
of the star is isothermal and the temperature gradients are
concentrated in the envelope.  The cooling rate of the star during
this period is determined by the processes of neutrino emission from
dense matter, whereby the neutrinos, once produced, leave the star
without further interactions. The understanding of the cooling
processes that take place during this neutrino radiation era is
crucial for the interpretation of the data on surface temperatures of
neutron stars. While the long term features of the thermal evolution
of neutron stars are insensitive to the non-isothermal cooling stage,
the subsequent route in the temperature versus time plane strongly
depends on the emissivity of matter during the neutrino-cooling
era. Thus, another outstanding question raised by the work of
Ref.~\cite{Demorset:2010} is how the massive neutron stars, featuring
quark matter, cool? The recent observation of the substantial change
in the temperature of the neutron star in Cas A poses a further
challenge for the theory to explain drastic short-term drop in the
temperature of this neutron
star~\cite{Heinke:2010cr,Shternin:2010qi}. Presently, consistent
calculations of the cooling of massive compact stars are virtually
absent. First steps in this direction have been taken by a number of
groups~\cite{Page:2000wt,Alford:2004zr,Anglani:2006br,arXiv:1104.1706}.

A robust feature of cold quark matter is its color
superconductivity~\cite{Alford:2007xm,Wang:2009xf}.  Unfortunately,
consistent, realistic simulations of cooling of compact stars
featuring quark matter are far from trivial because of the complex
phase structure of the quark matter at low temperatures and the
substantial effort that is needed to understand transport and weak
interactions in this matter.  The study of neutrino emissivities and
thermal conductivity of quark matter in the color superconducting
state is in its beginning. In some cases, even crude estimates are not
available for these quantities and a pressing requirements for any
realistic simulation of thermal evolution of massive compact stars is
the development of the knowledge of transport coefficients of these
phases.

\section{Two solar-mass compact stars with hyperons and color superconducting quarks}
\label{sec:sup_phases}

The existence of hybrid stars with two solar masses was predicted in a
number of models, including those based on the MIT model and NJL
models of quark matter and its superconductivity (for reviews see
Refs.~\cite{Alford:2007xm,Wang:2009xf}).  Our recent study based on
relativistic hypernuclear Lagrangians predicts stiff hypernuclear
EOS above saturation density~\cite{Bonanno:2011ch}.
This enables one to construct stable configurations with masses equal
and above the measured 1.97 solar-mass star. The resulting
configuration have ``exotic'' matter in their interiors in the form of
hyperons and quark matter, of which the quark matter is
color-superconducting in the two-flavor 2SC and/or three-flavor CFL
phases.

Here we briefly describe the set-up and main results of
Ref.~\cite{Bonanno:2011ch}.  The nuclear EOS, as is well
known, can be constructed starting from a number of principles, see,
\eg,
\cite{weber_book,Sedrakian:2006mq}. In
Ref.~\cite{Bonanno:2011ch} a number of relativistic mean-field models
were employed to model the low density nuclear matter.  As is
well-known, these models are fitted to the bulk properties of nuclear
matter and hypernuclear data to describe the baryonic octet and its
interactions~\cite{LBL-30645,206719}.  The underlying Lagrangian is
given by
\begin{eqnarray}
\label{eq:L_RMF}
\mathcal{L}_{B}&=&\sum_B \bar{\psi}_B[\gamma^{\mu}(i\partial_{\mu}-g_{\omega B}\omega_{\mu
}-\frac{1}{2} g_{\rho B} {\bm \tau} \cdot {\bm \rho_{\mu}} )
-(m_{B}-g_{\sigma B}\sigma)]\psi_B\nonumber\\
&+&\frac{1}{2}\partial^{\mu}\sigma\partial_{\mu
}\sigma-\frac{1}{2}m_{\sigma}^{2}\sigma^{2}
+\frac{1}{2}m_{\omega}^{2}\omega^{\mu}\omega_{\mu}-\frac{1}{4}{\bm
  \rho}^{\mu\nu} \cdot {\bm \rho}_{\mu\nu}+\frac{1}{2}m_{\rho}^{2}{\bm
  \rho}^{\mu}\cdot{\bm \rho}_{\mu}\nonumber\\
&-&\frac{1}{3}b m_{N} (g_{\sigma N} \sigma)^3
-\frac{1}{4}c (g_{\sigma N} \sigma)^4
+\sum_{e^-,\mu^-}\bar{\psi}_{\lambda}(i
\gamma^{\mu} \partial_{\mu}-m_{\lambda})\psi_{\lambda}
-\frac{1}{4}F^{\mu\nu}F_{\mu\nu}\text{,} \nonumber\\
\end{eqnarray}
where the $B$-sum is over the baryonic octet $B \equiv p, n, \Lambda,
\Sigma^{\pm,0}, \Xi^{-,0}$, $\psi_B$ are the corresponding Dirac
fields, whose interactions are mediated by the $\sigma$ scalar,
$\omega_{\mu}$ isoscalar-vector and $\rho_{\mu}$ isovector-vector
meson fields.  The next-to-last term in Eq.~(\ref{eq:L_RMF}) is the
Dirac Lagrangian of leptons, $F_{\mu\nu}$ is the energy and momentum
tensor of the electromagnetic field. The parameters in
Eq.~\eqref{eq:L_RMF} correspond to the NL3
parametrization~\cite{nucl-th/9607039}. Computations were made also
with the GM3 parameterization~\cite{LBL-30645}, with the result that
hyperonic matter cannot be accommodated within this model.  The choice
of this specific parametrization was made because the nucleonic matter
has the stiffest EOS compatible with the nuclear
phenomenology. The mean-field pressure of the (hyper)nuclear matter
can be obtained from Eq.~(\ref{eq:L_RMF}) in the standard
fashion~\cite{weber_book}.

The high-density quark matter was described in Ref.~\cite{Bonanno:2011ch}
by an  NJL Lagrangian, which
is extended to include the t' Hooft interaction term ($\propto K$) 
and the vector interactions ($\propto G_V$)~\cite{nucl-th/0602018}
\begin{eqnarray}
\label{eq:NJL_Lagrangian}
\mathcal{L}_{Q}&=&\bar\psi(i\gamma^{\mu}\partial_{\mu}-\hat m)\psi 
+G_V(\bar\psi i \gamma^{0}\psi)^2
+G_S \sum_{a=0}^8 [(\bar\psi\lambda_a\psi)^2+(\bar\psi i\gamma_5 \lambda_a\psi)^2]\nonumber\\
&+& G_D \sum_{\gamma,c}[\bar\psi_{\alpha}^a i \gamma_5
\epsilon^{\alpha\beta\gamma}\epsilon_{abc}(\psi_C)^b_{\beta}][(\bar\psi_C)^r_{\rho} 
i \gamma_5\epsilon^{\rho\sigma\gamma}\epsilon_{rsc}\psi^8_{\sigma}]\nonumber\\
&-&K \left \{ {\rm det}_{f}[\bar\psi(1+\gamma_5)\psi]+{\rm det}_{f}[\bar\psi(1-\gamma_5)\psi]\right\},
\end{eqnarray}
where the quark spinor fields $\psi_{\alpha}^a$ carry color $a = r, g,
b$ and flavor ($\alpha= u, d, s$) indices, the matrix of quark current
masses is given by $\hat m= {\rm diag}_f(m_u, m_d, m_s)$, $\lambda_a$
with $ a = 1,..., 8$ are the Gell-Mann matrices in the color space,
and $\lambda_0=(2/3) {\bf 1_f}$. Here $G_S$ and $G_D$ are the couplings 
in the scalar and di-quark channels.
The charge conjugate spinors are
defined as $\psi_C=C\bar\psi^T$ and $\bar\psi_C=\psi^T C$, where
$C=i\gamma^2\gamma^0$ is the charge conjugation matrix.  The partition
function of the system can be evaluated for the Lagrangian
(\ref{eq:NJL_Lagrangian}) neglecting the fluctuations beyond the
mean-field~\cite{nucl-th/0602018}.  To do so, one linearizes the
interaction term keeping the di-quark correlations $\Delta_c\propto
(\bar\psi_C)_{\alpha}^ai\gamma_5\epsilon^{\alpha\beta
  c}\epsilon_{abc}\psi_{\beta}^b$ and quark-anti-quark correlations
$\sigma_{\alpha}\propto\bar\psi_{\alpha}^a\psi_{\alpha}^a$. The pressure 
derived from \eqref{eq:NJL_Lagrangian} reads
\begin{eqnarray}
p&=&\frac{1}{2\pi^2}\sum_{i=1}^{18}\int_{0}^{\Lambda}dk k^2
\left\{\vert\epsilon_i\vert
+2T {\rm ln}\left[1
+\exp\left(-\frac{\vert\epsilon_i\vert}{T}\right)\right]\right\} 
+4 K \sigma_u\sigma_d\sigma_s
\nonumber\\
&-&\frac{1}{4G_D}\sum_{c=1}^{3}\vert\Delta_c\vert^2
-2G_s\sum_{\alpha=1}^{3}\sigma_{\alpha}^2+\frac{1}{4
  G_V}(2\omega_0^2+\phi_0^2)+\sum_{l=e^-,\mu^-}p_l-p_0-B^*,\nonumber\\
\end{eqnarray}
where $T$ is the temperature, $\epsilon_i$ are the quasiparticle
spectra of quarks, $\omega_0=G_V\langle QM \vert
\psi_u^{\dagger}\psi_u+\psi_d^{\dagger}\psi_d\vert QM\rangle$ and
$\phi_0=2 G_V\langle QM \vert \psi_s^{\dagger}\psi_s\vert QM\rangle$
are the mean field expectation values of the vector mesons $\omega$
and $\phi$ in quark matter, $p_l$ is lepton pressure, $p_0$ is the
vacuum pressure and $B^*$ is an effective bag constant.  The quark
chemical potentials are modified by the vector fields as follows
$\hat\mu^*={\rm diag}_f(\mu_u-\omega_0,\mu_d-\omega_0,\mu_s-\phi_0)$.
The numerical values of the parameters of the Lagrangians
\eqref{eq:L_RMF} and \eqref{eq:NJL_Lagrangian} are quoted in
Ref.~\cite{Bonanno:2011ch}.
\begin{figure}[tb]
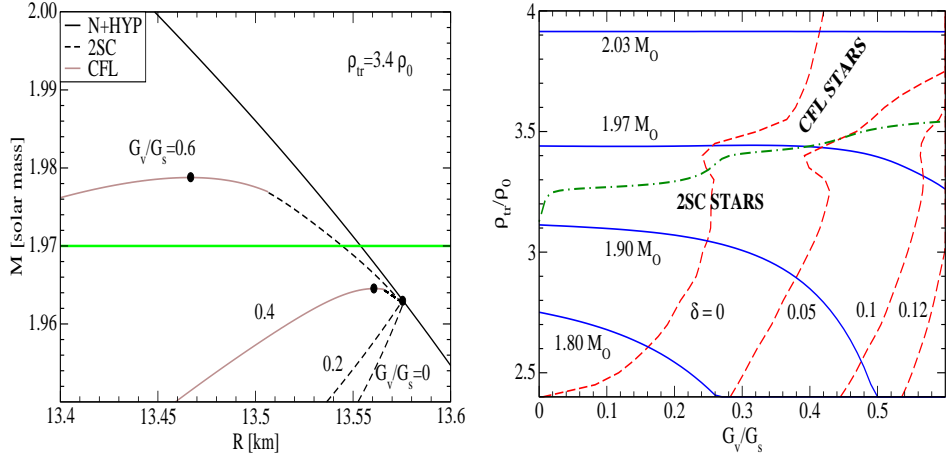

\hskip 0.1 cm
\begin{center}
\epsfig{figure=M_R,height=6.cm,width=6.cm,angle=0}
\hskip 0.2cm
\epsfig{figure=parameter_space2,height=6.cm,width=6.cm,angle=0}
\vskip 0.3cm
\caption{ 
{\it Left panel.}~Mass vs radius for configurations with quark-hadron
transition density $\rho_{\rm tr}=3.4\rho_0$ for four values of vector
coupling $G_V/G_S = 0, 0.2, 0.4, 0.6$. The purely hadronic sequence
(i.e. the sequence that includes nucleons and hyperons)
is shown by black solid line.  The dashed lines and the gray solid
lines show the branches where the 2SC and CFL quark phases are
present. The filled circles mark the maximum masses of the
sequences. The horizontal line shows the largest mass measurement to
date~\cite{Demorset:2010}.
{\it Right panel.}~
Properties of the stars as a function the free parameters
$G_V$ and $\rho_{\rm tr}$. The solid lines  show the
maximum mass configurations realized for the pair of parameters
$G_V$ and $\rho_{\rm tr}$. The dashed curves show the
amount of CFL matter in the configurations via the ratio $\delta =
R_{CFL}/R$, where $R_{CFL}$ is the radius of the CFL core, $R$ is
the star radius. The parameter space to the right from $\delta = 0$
line produces CFL stars.  The parameter space below the
dashed-dotted 5.1 curve corresponds to stars containing 
2SC matter.
}
\label{fig:1}
\end{center}
\end{figure}
The mass-radius relationship for massive stars constructed on the
basis of the EOSs described above is shown in the left panel of
Fig.~\ref{fig:1} together with the largest mass measurement to date
$M= 1.97\pm 0.04 M_{\sun}$ \cite{Demorset:2010}.  Masses above the
lower bound on the maximum mass are obtained for purely hadronic
stars; this feature is prerequisite for finding similar stars with
quark phases.  Evidently only for high values of vector coupling $G_V$
one finds stable stars that contain (at the bifurcation from the
hadronic sequence) the 2SC phase, which are followed by stars that
additionally contain the CFL phase (for higher central densities).
Thus, we find that the stable branch of the sequence contains stars
with quark matter in the 2SC and CFL phases.

The right panel of Fig.~\ref{fig:1} shows the changes in the masses
and composition of compact stars as the parameters of the model $G_V$
and $\rho_{\rm tr}$ are varied. First, it shows the tracks of constant
maximum mass compact stars within the parameter space. The decrease of
maximum masses with increasing vector coupling reflects the fact that
non-zero vector coupling stiffens the EOS. In other words,
to obtain a given maximum mass one can admit a small amount of soft
quark matter with vanishing vector coupling by choosing a high
transition density; the same result is obtained with a low transition
density, but strong vector coupling, \ie, a stiffer quark EOS. 
For low transition densities one finds 2SC matter in stars,
which means that weaker vector couplings slightly disfavor 2SC
matter. Substantial CFL cores appear in configurations for strong
vector coupling and almost independent of the transition density
(nearly vertical dashed lines with $\delta\sim 0.1$ in
Fig. \ref{fig:1}, right panel).  Note that for a high transition
density there is a direct transition from hyper-nuclear to the CFL
phase. For transition densities blow $3.5\rho_0$ a 2SC layer emerges
that separates these phases. On the other hand, weak vector couplings
and low transition densities produce stars with a 2SC phase only.

\section{Thermal evolution of massive stars }
\label{sec:equilibrium}

New features arise in the cooling behaviour of compact stars with the
onset of quark matter in sufficiently high-mass models. The recent
development of sequences of stable massive hybrid stars with realistic
input EOS~\cite{Ippolito:2007uz,Ippolito:2007hn,Knippel:2009st} allows
us to model the thermal evolution of compact stars containing quark
cores. These sequences of stable stars permit a transition from
hadronic to quark matter in massive stars ($M> 1.85 M_{\odot}$) with
the maximal mass of the sequence $\sim 2 M_{\odot}$.  In
Ref.~\cite{arXiv:1104.1706} quark matter of light $u$ and $d$ quarks was assumed
in beta equilibrium with electrons.  The pairing among the $u$ and $d$
quarks occurs in two channels: the red-green quarks are paired in a
condensate with gaps of the order of the electron chemical potential;
the blue quarks are paired with (smaller) gaps of order of keV, which
is comparable to core temperature during the neutrino-cooling
epoch~\cite{Alford:2002rz,Buballa:2002wy,Schmitt:2004et}.  For the
red-green condensate, a parameterization of neutrino emissivity was
chosen in terms of the gaplessness parameter $\zeta = \Delta/\delta\mu$, 
where $\Delta$ is the pairing gap in the red-green channel, $\delta\mu$ 
is the shift in the chemical potentials of the $u$ and $d$ 
quarks~\cite{Jaikumar:2005hy}.  The magnitude of the gap in the spectrum
of blue quarks was treated as a free parameter.  

The stellar models were evolved in time, with the input described
above, to obtain the temperature evolution of the isothermal interior.
The interior of a star becomes isothermal for timescales $t\ge
100$ yr, which are required to dissolve temperature gradients by
thermal conduction.  Unless the initial temperature of the core is
chosen too low, the cooling tracks exit the non-isothermal phase and
settle at a temperature predicted by the balance of the dominant
neutrino emission and the specific heat of the core {\it at the exit
  temperature}.  The
low-density envelope maintains substantial temperature gradients
throughout the entire evolution; the temperature drops by about 2
orders of magnitude within this envelope.  The iso\-thermal-interior
approximation relies further on the fact that the details of the
temperature gradients within the envelope are unimportant if we are
interested only in the surface temperature $T_s$. Models of the envelopes
predict the scaling $T_s^4 = g_s h(T)$, where $g_s$ is the surface
gravity, and $h$ is some function which depends on $T$, the opacity of
crustal material, and its EOS.  The fitted
formula $T_8 = 1.288 (T_{s6}^4/g_{s14})^{0.455}$~\cite{GPE:82} 
is commonly used.

In the isothermal-interior approximation, the parabolic differential
equation for the temperature reduces to an ordinary differential
equation, 
\be \label{eq:master} C_V \frac{dT}{dt} = -L_{\nu} (T)-L_{\gamma}(T_s)
+ H (T), \ee where $L_{\nu}$ and $L_{\gamma}$ are the neutrino and
photon luminosities, $C_V$ is the specific heat of the core, and the
heating processes, which could be important in the photon cooling era,
are neglected, \ie, $H(T) = 0$ (see
Refs.~\cite{Schaab:1999as,Gonzalez:2010ta} for a summary of these
processes and their effect on the evolution). 
\begin{figure}[hbt] 
\begin{center}
\includegraphics[width=12.0cm,height=8.0cm]{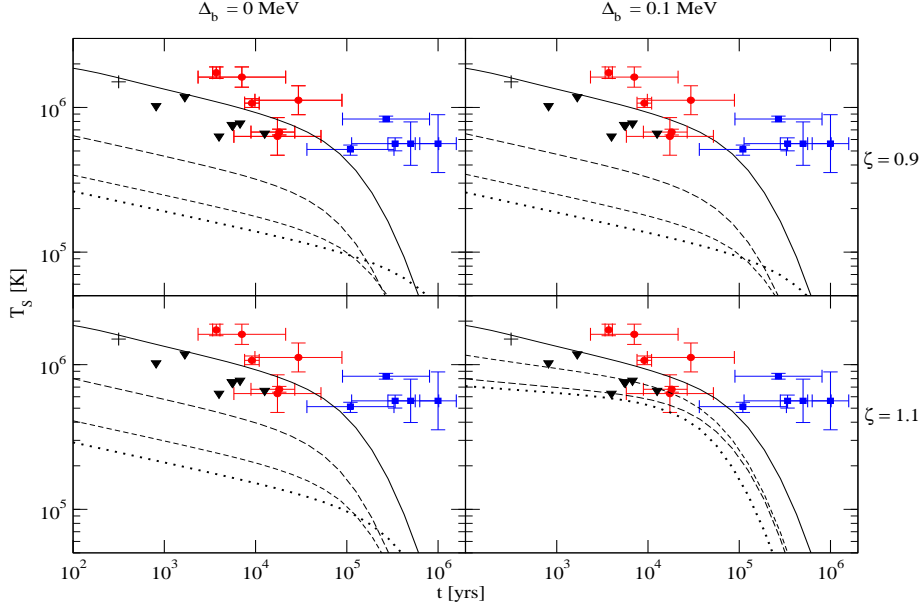}
\caption{ Time evolution of the surface temperature of four models
  with central densities   5.1
  (solid line), 10.8 (long-dashed line), 11.8 (short-dashed line),
  21.0 (dotted line) given in units of $10^{14}$ g cm$^{-3}$. 
   For the observational data see Ref.~\cite{arXiv:1104.1706}.
  The upper two panels correspond to cooling when the red-green
  condensate has $\zeta = 0.9$, i.e., is not fully gapped; the lower
  panels correspond to $\zeta = 1.1$, i.e., the red-green condensate is
  fully gapped. The left two panels correspond to evolution with
  negligible blue-quark pairing ($\Delta_b =0$); the right two panels
  show the evolution for large blue pairing $\Delta_b=0.1$~MeV.  }
\label{fig:3}
\end{center}
\end{figure}
The results of integration of Eq.~(\ref{eq:master}) are shown in
Fig.~\ref{fig:3}, where we display the dependence of the (redshifted)
surface temperature on time. Each panel of Fig.~\ref{fig:3} contains
cooling tracks for the same set of four models with central densities
5.1, 10.8, 11.8, 21.0 in units of $10^{14}$ g cm$^{-3}$. The
cooling tracks for the purely hadronic model (solid lines) are the
same in all four panels. The panels differ in the values of
micro-physics parameters, which characterize the pairing pattern in
quark matter.  Specifically, the two panels in the left column
correspond to the case where the blue-quark pairing is negligible
(\ie, the pairing is on a scale much smaller than the smallest energy
scale involved, typically the core temperature). The two panels in the
right column correspond to the case where the gap for blue quarks is
large, $\Delta_b = 0.1$ MeV. The panels in the upper and lower rows
are distinguished by the value of the $\zeta$ parameter.  [We use the
values $\zeta = 0.9$ (upper row) and $\zeta = 1.1$ (lower row)].  It
can be seen that (i) the neutrino-cooling is slow for hadronic stars
and becomes increasingly fast with an increase of the size of the
quark core, in those scenarios where there are unpaired quarks or
gapless excitations in the superconducting quark phase. The
temperature scatter of the cooling curves in the neutrino cooling era
is significant and can explain the observed variations in the surface
temperature data of same age neutron stars. (ii) If quarks of all
colors have gapped Fermi surfaces, the neutrino cooling shuts off
early, below the pairing temperature of blue quarks; in this case, the
temperature spread of the cooling curves is not as significant as in
the fast cooling scenarios. (iii) As the stars evolve into the photon
cooling stage the temperature distribution is inverted, \ie, those
stars that were cooler in the neutrino-cooling era are hotter during
the photon cooling stage.

\section{Perspectives}

The physics of massive compact stars poses a number of interrelated
questions/challenges. The first issue is the EOS of matter, including
the quark degrees of freedom and their color superconductivity.  The
strangeness degrees of freedom including hypernuclear matter and
three-flavor quark matter need to be further explored building, e.g.,
upon the work of Ref.~\cite{Bonanno:2011ch}. The equilibrium and
stability of massive compact objects, constructed from these EOSs,
should be studied including rapid rotations and oscillations.  Secondly,
we need a better understanding of the weak interaction rates in quark
and (hyper)nuclear matter, which are required input in cooling
simulations of compact stars. Thirdly, the transport coefficients of
dense color superconducting quark matter, such as the thermal
conductivity, are needed for modelling an array of phenomena, which
include thermal evolution, magnetic evolution, $r$-modes etc.

\section*{Acknowledgements}

I would like to thank Luca Bonanno and Daniel Hess for the collaboration
on the physics described in this lecture.  I
would like to thank Mark Alford, Xu-Guang Huang, Dirk H. Rischke and
Harmen Warringa for useful discussions.
 This work was in part supported by the Deutsche
Forschungsgemeinschaft (Grant SE 1836/1-2)

\end{document}